\documentclass{svjour3rev}                     
\smartqed  
\usepackage{graphicx}
\usepackage{bm}
\usepackage{amsmath,amsfonts,amsbsy,amssymb,makeidx, fullpage,array,graphicx,epsfig,enumerate}

\begin{document}

\title{Introduction to Bronstein's ``Quantum theory of weak gravitational fields"
}
\author{S. Deser and A. Starobinsky
}

\date{}

\maketitle

{
\vskip -2.5cm \noindent
S. Deser \\
Brandeis University, Waltham MA and Caltech, Pasadena CA, USA \\
e-mail deser@brandeis.edu \\
\\
Alexei A. Starobinsky \\
Landau Institute for Theoretical Physics RAS, Moscow, Russia \\
e-mail: alstar@landau.ac.ru
\vskip 1cm
}

\abstract{
A scientific introduction and short  biography to accompany the translation of  Matvei P. Bronstein, ``Quantum theory of weak gravitational fields",  {\it Phys. Zeitschr. der Sowjetunion} {\bf 9}, 140 157 (1936), to appear as a ``Golden Oldie" in JGRG.
}
\bigskip

Bronstein's is only the second work, after L.Rosenfeld's 1930 papers [1],
on the quantization of (linearized) gravity. It was carried out in 1935, at a
time when modern quantum field theory was in its infancy -- well before Fierz
and Pauli's treatment of spin 2, and soon after Dirac's (1931) and Fermi's
(1934) formulations of QED. Its birthplace -- Leningrad -- was also remarkable:
here a small group of young theoreticians were forming the nucleus of the future
Landau school and many other eminent Russian institutions. Indeed, Bronstein had
close ties, both personal and scientific, with Landau and his circle, and the
spirit animating this work clearly reflects these affinities of style,
recognizable to readers of the Landau \& Lifshitz series. Had Bronstein's life
not come to a tragic end only a few years later, in Stalin's purges, he would
certainly have fulfilled his promise as a most valuable contributor to forefront
theoretical physics.

The present work contains a, still quite readable and modern, exposition of
second quantization of the massless spin-2 field in flat space-time, including
even such essential details as the proper operator ordering required to avoid
zero-point energy. The system is quantized in Lorentz gauge, and shown to have
positive energy for each of its transverse-traceless graviton excitations. A
large part of the paper is devoted to the development of measurement theory for
a quantized gravitational theory -- the first attempt in this direction, too,
obviously influenced by the then-new Bohr - Rosenfeld analysis of quantum
electrodynamics. Though the author generally follows their treatment, he shows
that the gravitational field is essentially different due to the appearance of
another restriction on the accuracy of measurement of gravitational fields -- in
the case of large matter densities. [Here he concentrates on the Christoffel
symbols, as the first derivative equivalents of the EM fields.] From these
delicate and carefully treated physical effects, the author also concludes --
most wisely -- that it would be hardly possible to quantize full non-linear
gravity without deep revision of classical space-time notions -- prophetic
conclusion that remains very much actual!

Going beyond the free-field analysis, Bronstein further treated gravity-matter
interaction, concretely through Fock's novel covariantization of the Dirac
field. Here the author first considers energy transfer by quantized
gravitational waves and shows by lengthy calculation that, in the correspondence
limit of large occupation number, the classical Einstein expression follows
(after correcting a coefficient given by Einstein!). This shows that his
quantization is not only qualitatively, but also quantitatively correct.
Finally, following the, also contemporaneous, Fock - Podolsky derivation of
Coulomb's law from QED, he obtains Newton's law from his linearized quantum
gravity, carefully analyzing how the correct sign of the gravitational potential
-- i.e., opposite to that of Coulomb's -- appears in this formalism; while this
sign flip is already built-in at the classical level, it is a gratifying check
of the new quantum model's correctness.

Finally, we note again the elegant and thorough style and notation -- not only
for its time, but even now; as noted at the beginning, it will appeal, not
coincidentally, to anyone who has studied Landau - Lifshitz. In summary, this
paper certainly deserves revival, not just historically, but for its intrinsic
(and pedagogical) value. When it was published, general relativity had fallen
into a dormant phase, from which it only awoke two decades later. At last, three
quarters of a century since 1935, we begin the observational search for
quantum-gravitational effects in primordial gravitational wave background
generated during inflation. Published in German in a journal inaccessible today,
this accessible Golden Oldie translation is a welcome (re-)addition to our
subject!

{\bf Acknowledgement:} SD acknowledges support from NSF PHY-1064302 and DOE
DE-FG02-164 92ER40701 grants. We thank G. Gorelik and J. Stachel for very useful
correspondence.

 \bigskip

\bigskip


 \bigskip

\bigskip

{\bf Matvei P. Bronstein -- a brief biography}

\medskip

{\bf By S. Deser}

\medskip

Matvei Petrovich Bronstein was born December 2 1906, and shot in a Leningrad
prison in Stalin's purges on February 18, 1938 (rehabilitated posthumously in
1957). A precocious scientist, he was one of the legendary group that included
Landau and Gamow at Leningrad University in 1926. His first publications date
from 1925; in his one allotted decade, he covered a wide range of frontier
problems in cosmology, astrophysics and application of the then new quantum
mechanics. For his PhD thesis, a degree newly re-instituted in the USSR, he
chose the work presented here, which he defended in front of I.Tamm and V. Fock
in November 1935. In addition to his great scientific output and participation in
the surge of Soviet science in those early years, he translated Dirac's ``Quantum
mechanics", works by Bohr and wrote several children's science books. On August
6, 1937, he was arrested in the Great Purge; he was never seen again, despite
his widow's being informed that he was sentenced to ten years' prison. For a
detailed biography, see [1].

\end{document}